\documentclass{aa}
\usepackage{graphics}

\def\beq{\begin{equation}}
\def\eeq{\end{equation}}
\def\bey{\begin{eqnarray}}
\def\eey{\end{eqnarray}}
\def\kms{\rm \,km\,s^{-1}}

\def\kpc{\,{\rm {kpc}}}

\def\re{\tilde{r}_\oplus}
\def\tv{\tilde{v}}
\def\u0{u_0}
\def\tp{t_p}

\def\Ds{D_{\rm s}}
\def\Dd{D_{\rm d}}
\def\mI0{m_{\rm I,0}}
\def\mV0{m_{\rm V,0}}

\def\d{\thinspace{\rm d}}

\begin{document}

\thesaurus{01         
	(
	12.04.1;
	12.07.1;
	08.22.3 
	)}
\title{An Ongoing Parallax Microlensing Event 
OGLE-1999-CAR-1 Toward Carina}

\author{Shude Mao}

\offprints{S. Mao}
\mail{smao@mpa-garching.mpg.de}

\institute{Max-Planck-Institut f\"ur Astrophysik, 
	Karl-Schwarzschild-Strasse 1, 85740 Garching, Germany}

\date{Received 1999; accepted 1999}
\titlerunning{
An Ongoing OGLE Parallax Microlensing Event Toward Carina}
\authorrunning{Mao}

\maketitle

\begin{abstract}

We study the first microlensing event toward the Carina spiral arm
discovered by the OGLE collaboration. We demonstrate
that this long duration event exhibits strong parallax signatures.
Additional information from the parallax effect allows us to
determine the lens transverse
velocity projected onto the Sun-source line to be $\sim 145\kms$.
We also estimate the optical depth, event rate and duration
distribution for microlensing toward Carina.
We show that this event is broadly consistent with these predictions.

\end{abstract}

\section{Introduction}

Gravitational microlensing was originally proposed as a method of detecting
compact dark matter objects in the Galactic halo (Paczy\'nski
1986). However, it also turned out to be a powerful method to
study Galactic structure, mass functions of stars and extrasolar
planetary systems (for a review, see Paczy\'nski 1996).
Earlier microlensing targets include
the Galactic bulge, LMC, SMC and M31. Recently,
the EROS and OGLE collaborations have started to monitor spiral arms.
The EROS
collaboration (Derue et al. 1999) has announced the discovery 
of three microlensing events toward two spiral arms.
In this paper, we study the first spiral arm microlensing event
discovered by the OGLE II experiment (Udalski, Kubiak \& Szyma\'nski 1997).
This {\it ongoing} event, OGLE-1999-CAR-1, was discovered 
in real-time toward the 
Carina arm by the OGLE early-warning system (Udalski et al. 1994). 
We show that this is a unique
event that exhibits strong parallax effects. Such
events were predicted by Refsdal (1966) and Gould (1992). The first case was
reported by the MACHO collaboration toward the
Galactic bulge (Alcock et al. 1995).
OGLE-1999-CAR-1 is the first parallax event discovered toward any spiral arm.
The outline of the paper is as follows. In Sect. 2, we briefly describe
the observational data that we use.
In Sect. 3, we present two different fits for the light curve,
with and without considering the effect of parallax and blending.
In Sect. 4, we calculate the expected optical depth, event
rate and duration distribution toward Carina. Finally
in Sect. 5, we summarize and discuss the implications of our results.

\section{Observational Data}

The observational data were collected by the
OGLE collaboration in their second phase of microlensing search
(Udalski et al. 1997). The search was done with the 1.3-m Warsaw
telescope at the Las Campanas Observatory, Chile which is operated by
the Carnegie Institution of Washington. The targets of the OGLE II experiment
include the LMC, SMC, Galactic
bulge and spiral arms. All the collected data were reduced and calibrated
to the standard system. For more details on the 
instrumentation setup and data reduction, see
Udalski et al. (1997, 1998).

The event, OGLE-1999-CAR-1, was detected and announced in real-time
on Feb. 19, 1999 by the OGLE collaboration
\footnote{\hspace{-0.1cm}http://www.astrouw.edu.pl/\~\/ftp/ogle/ogle2/ews/ews.html}.
Its equatorial coordinates are
$\alpha$=11:07:26.72,
$\delta$=-61:22:30.6 (J2000), which corresponds to Galactic
coordinates $l=290^\circ.8, b=-0^\circ.98$.
The ecliptic
coordinates of the lensed star are
$\lambda=331^\circ.9$, $\beta=-58^\circ.1$ (e.g., Lang 1980).
Finding chart and 
I-band data of OGLE-1999-CAR-1 are available at the
above web site and the V-band data
were made available to us by Dr. Andrzej Udalski. 
In total, there are 416 I-band and 85 V-band data points.
The baseline magnitudes of the object 
\footnote{\hspace{-0.1cm} we added an offset of
$-0.003$ to the V-band and $+0.045$ to the I-band to obtain the standard
magnitudes (Udalski 1999, private communication)}, in the standard 
Johnson V and Cousins I-band, are $V=19.66, I=18.01$.

\section{Models}

In this section, we will fit the OGLE V and I-band light curves
simultaneously with theoretical models. We start with the simple
standard model, and then consider a fit that
takes into account both parallax and blending.

Most microlensing light curves are well described by the standard form
(e.g., Paczy\'nski 1986):
\beq \label{amp}
A(t) = {u^2+2 \over u \sqrt{u^2+4}},~~
u(t) \equiv \sqrt{\u0^2 + w^2(t)},
\eeq
where $\u0$ is the impact parameter (in units of the Einstein radius) and 
\beq
w(t) = {t-t_0 \over t_E}, ~~ t_E \equiv R_E/v_t
\eeq
with $t_0$ being the time of the closest approach (maximum
magnification), $R_E$ the Einstein radius, $v_t$
the lens transverse velocity relative to the observer-source line of
sight, and $t_E$ the Einstein radius crossing time. The Einstein radius
is defined as
\beq
R_E = \sqrt{4 G M \Ds x(1-x) \over c^2},
\eeq
where $M$ is the lens mass, $\Ds$ the distance to the source and
$x=\Dd/\Ds$ is the ratio of the distance to the lens and the distance
to the source.

To fit both the I-band and V-band data with the standard
model, we need five parameters, namely, 
\beq \label{pstandard}
\u0, t_0, t_E, \mI0, \mV0.
\eeq
Best-fit parameters are found by minimizing the usual $\chi^2$ 
using the MINUIT program in the
CERN library and are tabulated in Table 1. The resulting $\chi^2$ is
893.9 for 496 degrees of freedom. For convenience,
we divide the data into one `unlensed' part and one `lensed' part; the
former has $t={\rm J.D.}-2450000<1150\d$ and the latter has
$t>1150\d$. For the standard fit, the lensed part 
has $\chi^2=407.9$ for 161 data points, and the
unlensed part has $\chi^2=486.0$ for 340 data points; the somewhat
high $\chi^2$ for this part may be due to contaminations of nearby bright
stars (as can be seen in the finding chart), particularly 
at poor seeing conditions. The $\chi^2$
per degree of freedom for the `lensed' part is about 2.5, indicating
that the fit is not satisfactory.
This can also be seen from Fig. 1, where we
show the model light curve together with the data points. As can be seen,
the observed values are consistently brighter than the predicted ones
for $t>1325\d$ in the I-band. Further, the prediction is fainter by
about 0.05 magnitude at the peak in the V-band.
We show next that both inconsistencies can
be removed by incorporating parallax effect and blending.

\begin{figure*}
\begin{center}
\resizebox{0.85\hsize}{!}{\includegraphics{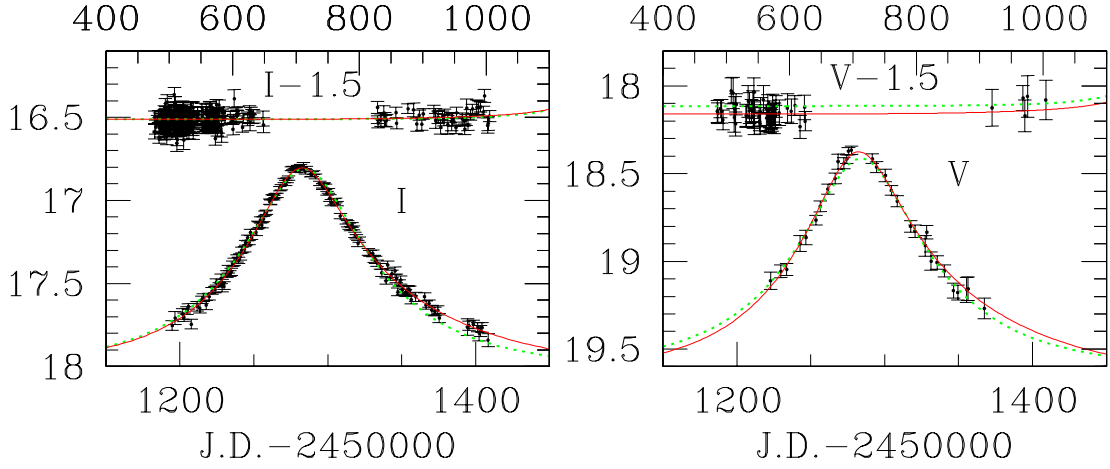}}
\end{center}
\vspace{-9.05cm}
\caption{
The I-band (left) and V-band (right) light curves observed by the OGLE 
collaboration are shown. In each panel, the constant
part of light curve is shown at the top with their amplitudes shifted by
1.5 magnitude and with their time intervals (from $400\d$ to $1100\d$)
labelled at
the top axis. The dotted line indicates the best-fit standard model
(eq. \ref{amp}), while the solid line is
for the best-fit model that takes into account both parallax
(eq. \ref{parallax}) and blending (eq. \ref{blend}). Fit parameters are
given in Table 1.
}
\end{figure*}

\begin{table*}
\caption[]{The best standard model (first row) and the best
parallax model with blending (second row) for OGLE-1999-CAR-1.
}
\begin{tabular}{lllllllllll}
\\
\hline\noalign{\smallskip}
& $t_0$ & $t_E$ & $\u0$ & $\mI0$ &  $\mV0$ & $\theta$ & $\tv$ &
$f_I$ & $f_V$ & $\chi^2$\\
\noalign{\smallskip}
\hline
\noalign{\smallskip}
\vspace{0.15cm}
S & $1284.2^{+0.1}_{-0.1}$ & $89.8^{+0.45}_{-0.45}$ 
	& $0.35^{+0.01}_{-0.01}$ & $18.01^{+0.02}_{-0.02}$
	& $19.62^{+0.05}_{-0.05}$ & --- & --- & --- & --- & 893.9 \\
\vspace{0.15cm}
P & $1288.8^{+1.7}_{-1.6}$ & $118.1^{+10.8}_{-6.9}$
	& $0.24^{+0.05}_{-0.03}$ &$18.01^{+0.02}_{-0.02}$
	& $19.66^{+0.09}_{-0.09}$&$-0.94^{+0.75}_{-0.73}$
	& $145.5^{+13.0}_{-40.8}$ & $0.36^{+0.2}_{-0.2}$
	& $0.29^{+0.2}_{-0.2}$ & 640.9 \\
\hline
\end{tabular}
\end{table*}


To take into account the Earth motion around the Sun, we have to modify
the expression for $u(t)$ in eq. (\ref{amp}). This modification, to the
first order of the Earth's orbital eccentricity ($\epsilon=0.017$), 
is given by Alcock et al. (1995) and Dominik (1998):
\bey \label{parallax}
u^2(t)& =& \u0^2+w(t)^2 + \re^2 \sin^2\psi 
	\nonumber \\
& + & 2\re \sin \psi \left[w(t) \sin\theta + \u0 \cos\theta \right]
+  \re^2\sin^2\beta \cos^2 \psi
	\nonumber \\ 
& + & 2\re \sin\beta \cos\psi \left[ w(t) \cos\theta - \u0 \sin\theta \right],
\eey
where $\theta$ is the angle between $\vec{v}_t$ and the line formed by
the north ecliptic axis projected onto the lens plane, $\u0$ is now
more appropriately the minimum distance between the lens and the
Sun-source line. The expression of $\re$ and $\psi$ are given by
\beq
\re = {1 {\rm AU} \over \tv~ t_E} \{1-\epsilon \cos[\Omega_0 (t-\tp)] \},
\eeq
and
\beq
\psi = -\phi + \Omega_0(t-\tp)+2\epsilon \sin[\Omega_0(t-\tp)],
\eeq
where $\tp$ is the time of perihelion, $\tv=v_t/(1-x)$ is the transverse
speed of the lens projected to the solar position, 
$\Omega_0=2\pi/{\rm yr}$, 
and $\phi$ is the longitude measured in the ecliptic
plane from the perihelion toward the Earth's motion; this 
is given in the appendix of Dominik (1998),
\beq
\phi=\lambda+\pi+\phi_\gamma,
\eeq
where $\phi_\gamma$
is the longitude of the vernal equinox measured from the perihelion.
$\phi_\gamma=1.33$ (rad), and the Julian day for Perihelion is
$\tp=2451181.57$; the readers are referred to the {\it The Astronomical
Almanac} (1999) for the relevant data.
Note that the inclusion of the parallax effect introduces two more 
parameters, $\tv$ and $\theta$. 

The two-color light curves show that the lensed object became
bluer by $\approx 0.05$ mag at the peak of magnification; such
chromaticity is easily produced by blending. The additional source of light
may be from the lens itself and/or it can come from another star
which lies in the seeing disk of the lensed star by chance alignment.
When blending is present, the observed magnification is given by
\beq \label{blend}
A_i = f_i + (1-f_i) A(t), ~~ i=I, V.
\eeq
To model the blending in two colors, we need two further
parameters -- the fraction of light contributed by the unlensed component
in I and V, $f_I$ and $f_V$, at the baseline. Therefore,
a fit that takes into account both parallax and blending
effects requires 9 parameters:
$\u0, t_0, t_E, \mI0$, $\mV0$, $\tv, \theta, f_I$, and $f_V$.

The best-fit parameters for this model are given in Table 1.
Compared with the standard fit, the
$\chi^2$ is reduced from 893.9 to 640.8. The reduction in 
the lensed part is dramatic: the $\chi^2$ 
drops from 407.9 to 177.7 for 161 data points. The $\chi^2$ for the
unlensed part is 463.2 (as compared to 486.0 for the standard fit) for
340 data points. The $\chi^2$ per
degree of freedom is satisfactory. 
The predicted light curve (solid line in Fig. 1)
matches the observed data both in the I-band and V-band. From Table 1,
the blending fractions in the V and I bands are
not well constrained, $f_I=0.36\pm 0.2$ and $f_V=0.29\pm 0.2$.
The differential blending, however, is reasonably constrained,
$f_V/f_I=0.81^{+0.1}_{-0.27}$
due to the observed differential magnification (0.05 mag)
between the I-band and V-band. The projected 
lens velocity is well constrained while 
its direction has somewhat larger errors.
For completeness, we mention that the best model that accounts for
blending but {\it not} parallax has $\chi^2=791.8$ while 
the best model that accounts for parallax but {\it not} blending
has $\chi^2=682.2$. 
Hence the parallax effect reduces the
$\chi^2$ much more effectively than blending. This can be
easily understood since the observed light curve is asymmetric,
which cannot be produced by blending.

\section{Optical Depth and Event Rate Toward Carina}

To put OGLE-1999-CAR-1 into context, in this section we estimate the
optical depth, event rate and event duration distribution
toward Carina. These can be compared with future observations when
more events become available.

One major uncertainty for microlensing toward spiral arms such as Carina
is that we do not know the distance to the sources.
Several molecular clouds at the same direction have distances
between $6.8\kpc$ and 
$8.7\kpc$ (see Table 1 and Fig. 4. in Grabelsky et al. 1988).
We adopt $\Ds=6.8\kpc$ as the distance to the sources. This
assumption is also consistent with the lensed star being a main-sequence
star as required by its position
in the color-magnitude diagram (Udalski
1999, private communication); we return to this issue in the discussion.

Since the direction toward the lensed star is nearly in the Galactic
plane ($b=-0^\circ.98$)
and far away from the Galactic center, we assume that the lenses
are entirely contributed by disk stars, and their density profile follows
the standard double exponential disk distribution
\beq
\rho(R, z) = {\Sigma \over 2 z_0} ~ \exp\left(-{r-r_0 \over r_d}\right) 
\exp\left(-{|z| \over z_0}\right),
\eeq
where $\Sigma=50 M_\odot/{\rm pc}^2$, 
$r$ is the lens distance to the Galactic center, $r_0=8.5\kpc$,
disk scale-length $r_d=3.5\kpc$ and scale-height $z_0=0.325\kpc$.
The optical depth can then be obtained (e.g., eq. 9 in
Paczy\'nski 1986):
\beq
\tau = 3.4 \times 10^{-7},
\eeq
independent of the lens mass function and kinematics.

\begin{figure}
\resizebox{\hsize}{!}{\includegraphics{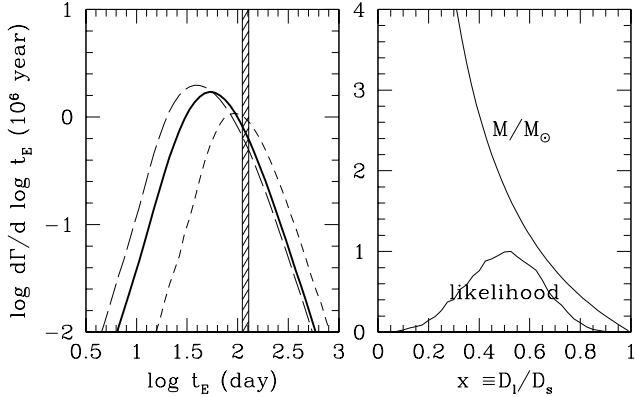}}
\vspace{-3.5cm}
\caption{
The left panel shows the predicted event rate distribution (in units of per 
million stars per year) as a function of duration toward Carina.
The short dashed line
is for a $\delta$-mass function of $1M_\odot$, the long dashed line
is for a Salpeter mass function, $n(M) dM \propto M^{-2.35} dM$ and the
solid line is for a disk mass function,
$n(M) dM \propto M^{-0.54} dM$, determined from HST star counts.
The shaded region indicates the $1\sigma$ range in time-scale for the best
parallax fit with blending. The right panel shows the likelihood
function and lens mass as a function of the lens distance, respectively.
}
\end{figure}

To estimate the event rate and event duration distribution, we have to
make assumptions about the lens kinematics and mass function. 
The motion of lenses can be divided into an
overall Galactic rotation of $220\kms$ 
and a random motion. We assume that the random component follows Gaussian
distributions with velocity dispersions of
$\sigma_R=40\kms, \sigma_\theta=30\kms, \sigma_z=20\kms$
(cf. Derue et al. 1999). The motion of
the Sun relative to the Local Standard of Rest is taken as
$v_{R,\odot}=9\kms, v_{\theta,\odot}=11\kms, 
v_{z,\odot}=16\kms$. We study three mass functions:  \par
1. $n(M) dM \propto \delta(M-1 M_\odot) dM$, \par
\vspace{0.05cm}
2. $n(M) dM \propto M^{-2.35} dM,~~ 0.08 M_\odot < M < 1 M_\odot$, \par
\vspace{0.05cm}
3. $n(M) dM \propto M^{-0.54} dM,~~ 0.1 M_\odot < M < 0.6 M_\odot$. \par
Note that the second is a Salpeter mass function while the 
third describes the local disk mass function determined
using HST star counts (Gould et al. 1997).
The predicted event duration distributions for these three mass functions
are shown in the left panel of Fig. 2. It is clear that there is a tail
toward long durations for all three distributions. The
probabilities of having $t_E$ longer than $100\d$ are respectively,
$\sim$ 50\%, 10\% and 20\%; so the observed long duration is not 
statistically rare.
The total event rates per million stars per year are found to be
\beq
\Gamma=0.64, 1.4, 1.1,
\eeq
for the three mass functions, respectively. The predicted duration
distribution and event rate are sensitive to the 
assumed velocity dispersions. For example, if we adopt
$\sigma_R=\sigma_\theta=\sigma_z=20\kms$ (as in Kiraga \& Paczy\'nski
1994), then the events will be on average
30\% longer and the event rate decreases by about 30\%.

\section{Discussion}

We have shown that the microlensing event, OGLE-1999-CAR-1,
has a light curve shape that is significantly modified 
by the earth motion around the Sun. 
This event is still {\it ongoing} at the time
of writing (Aug. 12, 1999); later evolutions of the light curve
will test our predictions and reduce the uncertainties in the parameters.
For comparison, the three
microlensing events seen by the EROS collaboration (Derue et al. 1999)
are toward different spiral arms which are somewhat 
closer to the Galactic center direction. The three events have $t_E$
between 70$\d$ to 100$\d$, and none of the events show parallax effects.

We have assumed a source distance of $6.8\kpc$ in the previous
section. We show now that this is a reasonable assumption. From
$V=19.66$ and $I=18.01$, and
taking into account the blending, we find that
the lensed star has $M_V=5.9, M_I=4.3$. Assuming an extinction of
$A_V=1.5$ mag (see Fig. 4 in Wramdemark 1980), and $A_I/A_V=0.482$, 
we obtain the intrinsic magnitude and color $M_V=4.4,
(V-I)_0=0.75$. The star is consistent with being a main sequence star (with
mass $M \sim 1.05 M_\odot$),
as can be seen from the color-magnitude diagram (Udalski 1999, private
communication) in the Carina region.


We can combine the expression for
$\tv$ and $t_E$ to obtain the lens mass as a function of its distance
\beq
M = {1-x \over x} {\tv^2 t_E^2 c^2 \over 4 G \Ds}
\approx 1.8 M_\odot {1-x \over x}.
\eeq
Using eq. (6) in Alcock et al. (1995), we have calculated the likelihood of
obtaining the observed transverse velocity and direction as a function
of the lens distance. The result is shown in the right panel of Fig. 2
for the Gould et al. (1997) mass function.
The lens distance is approximately $x=0.5\pm 0.2$; we caution that this
likelihood function is sensitive to the assumed kinematics. For example,
if we take $\sigma_R=\sigma_\theta=\sigma_z=20\kms$, then 
the best-fit lens distance changes to $x=0.7$.
It is possible that a lens is a (massive) dark lens
such as a white dwarf, neutron star and black hole; in this case
the blended light is contributed by another unrelated star.
Another possibility 
is that the blending is contributed by the lens itself. If the lens is a
main sequence star and $x \approx 0.7$, which implies $M\approx 0.8
M_\odot$, then the lens will contribute about the right amount of 
light to explain the light curves (including the
color change in the V and I bands). Further high-resolution imaging 
of the lensed object should allow us to differentiate between
these two possibilities.

\begin{acknowledgements}
We greatly appreciate the OGLE collaboration for 
permission to use their data for this publication. I also thank
Matthias Bartelmann, Martin Dominik,
Bohdan Paczy\'nski, Andrzej Udalski, Hans Witt and particularly
the referee Eric Aubourg for 
valuable discussions and comments on the paper.
\end{acknowledgements}

\end{document}